\documentclass[twocolumn]{aastex61}

\received{August 17, 2017}
\accepted{July 20, 2018}
\submitjournal{ApJ}

\shorttitle{Optical and Radio Polarizations in OJ~287}
\shortauthors{Sasada et al.}

\begin{document}

\title{Optical Emission and Particle Acceleration in a Quasi-Stationary Component in the Jet of OJ~287}

\correspondingauthor{Mahito Sasada}
\email{sasadam@hiroshima-u.ac.jp}

\author[0000-0001-5946-9960]{Mahito Sasada}
\affil{Hiroshima Astrophysical Science Center, Hiroshima University, 1-3-1 Kagamiyama, Higashi-Hiroshima, Hiroshima 739-8526, Japan}
\affil{Mizusawa VLBI Observatory, NAOJ, 2-12 Hoshigaoka, Mizusawa-ku, Oshu, Iwate 023-0861, Japan}
\affiliation{Institute for Astrophysical Research, Boston
University, 725 Commonwealth Avenue, Boston, MA 02215, USA}

\author{Svetlana Jorstad}
\affiliation{Institute for Astrophysical Research, Boston
University, 725 Commonwealth Avenue, Boston, MA 02215, USA}
\affiliation{Astronomical Institute, St. Petersburg University, Universitetskij Pr. 28, Petrodvorets, 198504, St. Petersburg, Russia}

\author{Alan P. Marscher}
\affiliation{Institute for Astrophysical Research, Boston
University, 725 Commonwealth Avenue, Boston, MA 02215, USA}

\author{Vishal Bala}
\affiliation{Institute for Astrophysical Research, Boston
University, 725 Commonwealth Avenue, Boston, MA 02215, USA}

\author{Manasvita Joshi}
\affiliation{Institute for Astrophysical Research, Boston
University, 725 Commonwealth Avenue, Boston, MA 02215, USA}

\author{Nicholas R. MacDonald}
\affiliation{Institute for Astrophysical Research, Boston
University, 725 Commonwealth Avenue, Boston, MA 02215, USA}
\affiliation{Max-Planck-Institute f\"ur Radioastronomie, Auf dem H\"ugel 69, D-53121 Bonn, Germany}

\author{Michael P. Malmrose}
\affiliation{Institute for Astrophysical Research, Boston
University, 725 Commonwealth Avenue, Boston, MA 02215, USA}

\author{Valeri M. Larionov}
\affiliation{Astronomical Institute, St. Petersburg University, Universitetskij Pr. 28, Petrodvorets, 198504, St. Petersburg, Russia}

\author{Daria A. Morozova}
\affiliation{Astronomical Institute, St. Petersburg University, Universitetskij Pr. 28, Petrodvorets, 198504, St. Petersburg, Russia}

\author{Ivan S. Troitsky}
\affiliation{Astronomical Institute, St. Petersburg University, Universitetskij Pr. 28, Petrodvorets, 198504, St. Petersburg, Russia}

\author{Iv\'an Agudo}
\affiliation{Instituto de Astrof\'{\i}sica de Andaluc\'{\i}a, CSIC, Apartado 3004, E-18080, Granada, Spain}

\author{Carolina Casadio}
\affiliation{Instituto de Astrof\'{\i}sica de Andaluc\'{\i}a, CSIC, Apartado 3004, E-18080, Granada, Spain}
\affiliation{Max-Planck-Institute f\"ur Radioastronomie, Auf dem H\"ugel 69, D-53121 Bonn, Germany}

\author{Jos\'e L. G\'omez}
\affiliation{Instituto de Astrof\'{\i}sica de Andaluc\'{\i}a, CSIC, Apartado 3004, E-18080, Granada, Spain}

\author{Sol N. Molina}
\affiliation{Instituto de Astrof\'{\i}sica de Andaluc\'{\i}a, CSIC, Apartado 3004, E-18080, Granada, Spain}

\author{Ryosuke Itoh}
\affiliation{Department of Physics, School of Science, Tokyo Institute of Technology, 2-12-1 Ohokayama, Meguro, Tokyo 152-8551, Japan}



\begin{abstract}

We analyze the linear polarization of the relativistic jet in BL Lacertae object 
OJ~287 as revealed by multi-epoch Very Long Baseline Array (VLBA) images at 43 
GHz and monitoring observations at optical bands. The 
electric-vector position angle (EVPA) of the optical polarization matches that at 43 
GHz at locations that are often in the compact millimeter-wave ``core'' or, at
other epochs, coincident with a bright, quasi-stationary emission feature $\sim0.2$~milliarcsec 
($\sim$0.9~pc projected on the sky) downstream from the core. This implies that electrons
with high enough energies to emit optical synchrotron and $\gamma$-ray inverse Compton
radiation are accelerated both in the core and at the downstream feature, the latter of 
which lies $\geq10$~pc from the central engine. The polarization
vector in the stationary feature is nearly parallel
to the jet axis, as expected for a conical standing shock capable of accelerating
electrons to GeV energies.

\end{abstract}

\keywords{galaxies: active - galaxies: jets - BL Lacertae objects: individual (OJ 287) - techniques: polarimetric - methods: observational}



\section{Introduction}
Production of the optical synchrotron radiation observed in blazars requires 
electrons with energies $\gtrsim 3$~GeV. Similar electron energies are 
needed for inverse Compton scattering to generate the GeV $\gamma$-ray 
emission often detected in the same objects 
\citep[e.g.,][]{2016Galax...4...37M}. There are multiple possible sites in blazar
jets where particles might be accelerated to such 
energies on (sub)parsec scales: (1) near the base of the jet, 
$\sim 20$ gravitational radii from the black hole \citep{2011Natur.477..185H}, (2) in 
a magnetically dominated acceleration/collimation zone (ACZ) 
\citep{2008Natur.452..966M,2010ApJ...710L.126M,2015MNRAS.450..183S}, 
(3) in regions where the plasma is turbulent, (4) in parsec-scale standing shocks
\citep{2010ApJ...715..355L,2014ApJ...780...87M}, and (5) in moving shocks that propagate 
from the jet base, through the parsec-scale ``core'' and beyond 
\citep[e.g.,][]{1985ApJ...298..114M,2011ApJ...727...21J}. Each of these 
possibilities corresponds to a distinct magnetic field geometry: helical field at 
the jet base, anti-parallel field lines in magnetic reconnection zones, 
disordered field lines in turbulent regions, and field parallel to shock fronts.
Observations of the linear polarization of synchrotron radiation are sensitive 
probes of the magnetic field geometry, and therefore can test the various
possibilities for particle acceleration.

In this study, we probe the magnetic field geometry in the millimeter-wave and optical
emission zones of the BL Lacertae object OJ~287 (redshift $z=0.306$, at which
1~milliarcsecond (mas)=4.5~pc; 
\citealt{2010A&A...516A..60N}) by combining multi-epoch total and 
polarized intensity Very Long Baseline Array (VLBA) images at 43~GHz with 
well-sampled optical monitoring of the linear polarization. The VLBA images, with an
angular resolution of $\sim0.15$~mas,  
resolve the innermost regions of the jet. Comparison with the optical polarization allows 
us to determine the most likely site of optical synchrotron emission in the jet, 
as well as the geometry of the magnetic field in that location, at different times.

The compact jet of OJ~287 has changed dramatically over time 
\citep{2012ApJ...747...63A}. Prior to 2005, it was directed to the southwest 
\citep{2005AJ....130.1418J}, in a similar direction as the arcsecond-scale radio/X-ray 
jet \citep{2011ApJ...729...26M}. Its electric-vector position angle (EVPA) then abruptly 
switched by $\gtrsim90^{\circ}$ to the north and then northwest 
\citep{2011ApJ...735L..10A}. Subsequently, a bright, quasi-stationary feature, ``$S$'',
developed about 0.2~mas northwest of the bright compact ``core'' situated at the upstream
end of the jet in 43 GHz VLBA images\citep{2011ApJ...735L..10A,2017A&A...597A..80H}. 

As is common for low-frequency-peaked BL Lac objects like OJ~287, the 
linear polarization at both millimeter and optical wavelengths is highly variable. 
For example, \citet{2009ApJ...697..985D} measured changes from $9\%$ to 
$15\%$ in the degree of polarization of the core in 7~mm VLBA images, and 
from $8\%$ to $30\%$ at optical $R$-band during the same time span. 
Similarly, \citet{2010MNRAS.402.2087V} reported optical polarization
variations over a range from $\sim0$ to $35\%$ over 4.5~yr. This behavior
provides the potential to locate the optically emitting region on VLBA
images by matching the optical EVPA with that of a millimeter-wave feature
\citep{2016Galax...4...37M}. Here we apply this principle to OJ~287 over a period of nine
years.

\section{Observations and Data Analysis}
\subsection{VLBA Imaging and Optical Polarimetry}

OJ~287 is one of 37 blazars monitored roughly monthly at a wavelength of 7~mm
with the VLBA under the VLBA-BU-BLAZAR 
project\footnote{http://www.bu.edu/blazars/VLBAproject.html} 
\citep{2016Galax...4...47J}. Calibration of the data, as well as total and polarized 
intensity imaging procedures, are essentially identical to those described by 
\citet{jor17} and \citet{2007AJ....134..799J}. In order to calibrate 
the absolute value of the EVPA of the VLBA data, we use (1) Very Large Array 
observations at 7~mm of a number of VLBA-BU-BLAZAR blazars, including OJ~287,
within 1-3 days of the VLBA observations; and (2) the ``D-terms'' method 
\citep{2001ApJ...561L.161G}. We analyze Stokes-parameter $I$, $Q$, and $U$ 
images at 95 epochs from 2007 June~14 to 2016 March~18. Time intervals of VLBA 
observations are roughly 1 month, and most of them are within 2 months. 
All images are convolved with an elliptical Gaussian restoring beam of FWHM
0.33$\times$0.16~mas oriented along position angle PA$=-$10$^\circ$, 
approximating the angular resolution with uniform weighting of the 
visibilities.
To check for signs of Faraday rotation and optical depth effects, we compared the EVPA in
the 43~GHz images when the EVPA was uniform across the main emission
regions with a measurement of the integrated 86 GHz polarization at the 30~m antenna of
the Institut de Radioastronomie Millim\'etrique \citep{agudo17a,agudo17b} within 2 days of
the VLBA observation. Six epochs from 2008.50 to 2014.57 met these criteria. The mean
43-86 GHz EVPA difference was $2^\circ\pm3^\circ$, leading us to conclude that the 43~GHz
polarization was unaffected by either Faraday rotation or synchrotron self-absorption.

We use data from optical polarization observations of OJ~287 performed at:
(1) the 1.83~m Perkins 
Telescope of Lowell Observatory (Flagstaff, AZ) in $R$-band with the PRISM 
camera\footnote{http://www.bu.edu/prism}; (2) the 0.7~m AZT-8 telescope of the 
Crimean Astrophysical Observatory (Nauchnij, Crimea) in $R$-band; (3) the 0.4~m LX-200 
telescope of St. Petersburg University (St. Petersburg, Russia) without filter 
(central wavelength ${\lambda}_{\rm eff}\approx$670~nm); (4) the 1.54~m 
Kuiper and 2.3~m Bok telescopes of Steward Observatory (Tucson, AZ) in 
spectral-polarimetric mode covering a wavelength range of 
4000--7550~\AA\footnote{http://james.as.arizona.edu/$\sim$psmith/Fermi/}; (5) the 2.2~m 
Telescope of Calar Alto Observatory (Almer\'ia, Spain) in $R$-band with the
CAFOS multi-purpose focal reducer (acquired as part of the MAPCAT 
project\footnote{http://www.iaa.es/$\sim$iagudo/\_iagudo/MAPCAT.html}), and (6) the 
1.5~m Kanata telescope of Higashi-Hiroshima Observatory (Hiroshima, Japan) in 
$V$-band. Details of the polarimetric data reduction can be found in 
\cite{2010ApJ...715..362J,2008A&A...492..389L,2009arXiv0912.3621S}; and 
\cite{2011PASJ...63..639I}. The differences in EVPA among the different optical bands 
are negligible compared with the temporal variations during the time spans between the
observations. 
Finally, most of time intervals are within 1-d in the time series of measured polarization.

We have examined the data sets to identify 43 pairs of VLBA and 
optical polarization observations simultaneous within $\pm$1~d. The optical 
polarization measurements were collected nearly every night during periods when 
OJ~287 was visible. At epochs when multiple optical measurements matched the 
simultaneity criterion, we average the $I$, $Q$, and $U$ values to obtain the degree
$P_{\rm opt}=\sqrt[]{Q^{2}+U^{2}}/I$ and position angle 
$\chi_{\rm opt}=\frac{1}{2}\;{\rm tan}^{-1}\left(\frac{U}{Q}\right)$.

We employ the {\it Gnu R} programming language (v3.2.1)\citep{RManual} for 
analysis of the VLBA images at these 43 epochs. At each epoch we 
determine $I_{\rm i}$, $Q_{\rm i}$, and $U_{\rm i}$ as the mean values over 
circular area $\rm i$, ${\rm i}=1$ to $N$; the circular areas have radii of 0.05~mas 
and centers spaced every 0.05~mas along the jet trajectory. The averaged $N$ of 
43 epochs is approximately 25. The degree of 
polarization $P_{\rm i}$ and EVPA $\chi_{\rm i}$ of each area are calculated from 
the corresponding Stokes parameters. The uncertainties ${\delta}I$, ${\delta}Q$, 
and ${\delta}U$ are derived as the mean values of the noise in eight 
off-source regions in each $I$, $Q$, and $U$ image. The uncertainties of 
$P_{\rm i}$ and $\chi_{\rm i}$ are calculated via standard propagation of errors
\citep[e.g.,][]{1999PASP..111..898K}.
The results are corrected for 
Rice bias as described by \citet{1974ApJ...194..249W}, with data for which 
$P_{\rm i}/{\delta}P_{\rm i}<0.5$ excluded from consideration.

\subsection{Determination of Jet Trajectory}

Optically thin synchrotron radiation has only a weak wavelength dependence 
of polarization, even as the spectral slope changes. The optically emitting 
region can, therefore, potentially be identified on VLBA images by comparing optical
with radio polarization along the jet, provided that optical depth and 
Faraday effects are weak, as in OJ~287. In order to do this, we determine the jet
trajectory in each VLBA image and calculate Stokes parameters $I$, $Q$, $U$ to 
determine the polarization vector at each step along the trajectory.

\begin{figure*}
\begin{center}
\begin{tabular}{c}
\includegraphics[scale=0.6]{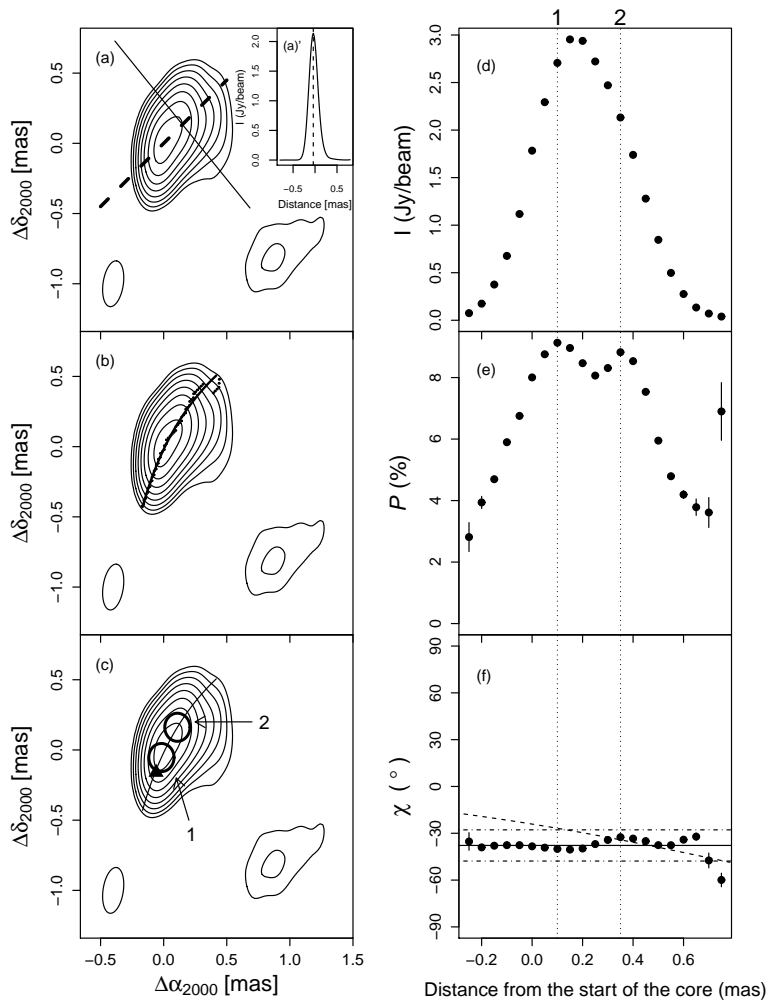}
\end{tabular}
  \caption{Example of VLBA images at 43~GHz and profiles of intensity and 
  polarization along the OJ~287 jet obtained on 2010 November 1. Restoring 
  beam corresponds to ellipse in lower left corners of panels (a)-(c). 
  Contour levels represent 0.5, 1, 2, 4, 8, 16, 32, and 64\% of the peak 
  intensity of 3.13 Jy$/$beam. Dashed line in panel~(a) delineates rough 
  jet direction. Intensity profile along the solid line is shown in 
  panel~(a)$'$ as an example of how trajectory points are determined (see 
  text). Black circles in panel~(b): maximum-intensity position at each 
  radial distance; solid curve; best functional fit. 
  Black triangle in panel~(c): defined upstream boundary of the core.
  Panel (c) divides main emission structure into two circular Gaussian components 
  with radii of 0.1~mas whose FWHM contours are indicated: (1) ``core'' and 
  (2) stationary feature ``$S$,'' with centers indicated by dotted lines in 
  panels (d)-(f). Panels (d)-(f) show profiles of intensity, polarization, 
  and EVPA along jet trajectory. Solid, dash dot, and dashed lines in panel~(f) 
  indicate, respectively, optical EVPA, $\pm$10$^{\circ}$ areas of optical 
  EVPA, and PA of jet trajectory. 
  }
  \label{fig:Nov}
\end{center} 
\end{figure*}

We perform the following procedures to determine the jet trajectory. 
(1) A line is drawn that roughly follows the jet axis on the VLBA 
image. (2) Line segments extending to $\pm$0.8~mas from the axis are 
placed perpendicular to the local jet direction. The dashed and solid 
lines in panel (a) of Figure~\ref{fig:Nov} illustrate this procedure. 
(3) The maximum intensity position is located on the solid line shown 
in panel (a)$'$ of Figure~\ref{fig:Nov}, which displays the intensity 
profile along the solid line. (4) The maximum intensity positions are 
determined every 0.01 or 0.02~mas, depending on the pixel size of the 
image. (5) A polynomial function corresponding to the jet trajectory 
is fit to these maximum  intensity positions, as illustrated in 
panel~(b) of Figure~\ref{fig:Nov}. We define the upstream boundary of 
the ``core'' of the jet as the position of the maximum gradient in the 
intensity profile along this trajectory.

The variation timescale of this object at optical wavelengths, 
${\delta}{\rm t}$, is $\sim0.8$~d, as determined by the {\it Kepler} 
spacecraft \citep{2017arXiv170904457G}. Setting this to the
light-crossing time allows us to estimate the size of the high-energy
emitting zone as ${\delta}{\rm t}cD/\left(1+z\right)\sim$0.0046~pc,
where $c$ is the speed of light and $D\sim9$ is the Doppler factor
\citep{jor17}. The 
spatial distance corresponding to the angular resolution of the VLBA 
image of 0.1~mas is $\sim1.4$~pc 
($=\theta_{\rm VLBA}/{\rm sin}{\Theta}$, where the angle between the jet
axis and our line of sight, $\Theta$, is assumed to be 3$^{\circ}$). The 
spatial distance is much larger than the size of the emitting zone estimated 
from the timescale of optical variations. Several optical emitting zones
could therefore be contained within the spatial resolution of the VLBA. 

We calculate the average $I$, $P$, and $\chi$ values within each 
circular region of radius 0.05~mas which is a half of the spatial 
resolution along the jet trajectory. 
Panels~(d)-(f) of Figure~\ref{fig:Nov} show an example of the 
distributions of the calculated intensity and polarization parameters. 
In this example, $\chi$ at 7~mm agrees well with that at optical 
wavelengths (solid line). 
The image and double-peaked polarization profile in Figure~\ref{fig:Nov}
reveal the presence of a component, ``$S$,'' northwest of the core, as 
also found by \citet{2011ApJ...735L..10A} and \citet{2017A&A...597A..80H}. 

\subsection{Definition of Coincidence between the Optical and Radio EVPAs}

If the optical and radio emissions are co-spatial, then the EVPA of the radio 
emission should be similar to the optical EVPA. However, there is also the
possibility of a chance coincidence between the varying radio EVPA 
in a more extended radio-emitting region and the observed optical EVPA.
To assess the extent to which such spurious coincidences affect our results,
we evaluate the systematic uncertainty of a chance coincidence between a 
random radio EVPA and the optical value. We generate 1000 uniform random 
numbers representing the radio EVPA ranging from 0$^{\circ}$ to 180$^{\circ}$. 
These numbers are subtracted from 1000 other random numbers selected from a 
normal distribution having the same mean as the observed optical EVPAs and a 
standard deviation equal to its uncertainty. The distribution of 
these differences should be related to the probability of chance 
coincidence between the optical and random radio EVPAs. We also estimate 
the distribution of differences between the random numbers selected from normal 
distributions having the same means as the observed optical and radio EVPAs.
We define EVPA-coinciding positions as those where the differences between the 
radio and optical EVPAs are less than 10$^{\circ}$, which is the 
standard deviation of random fluctuations in the optical EVPA over 
100-200 day intervals when there are no systematic EVPA variations. 
If the number of differences within 10$^{\circ}$ between the observed optical and 
radio EVPAs is at least three times larger than those between simulated EVPAs 
selected from the normal (optical) and uniform (radio) distributions, corresponding 
to 3 sigma confidence, we classify the coincidence between the observed optical and 
radio EVPAs as significant. 
We evaluate the coincidence between the radio 
EVPAs at each position along the jet trajectory and optical EVPAs 
in this manner.

Positions of low ($<100$) polarized intensity signal-to-noise ratios are 
excluded from the analysis. Applying this procedure to all 43 epochs, we 
find at least one EVPA-coinciding position along the jet trajectory in 36
of the epochs. 

\section{Results}
\subsection{Correlation between Radio and Optical Polarizations}

In the appendix we show all of the estimated profiles of the intensity, degree of 
polarization, and EVPA along the jet trajectory. We determine approximate values of 
these quantities for the core and stationary feature $S$ from the profiles. 
The profiles varied in time, as shown in Figures \ref{fig:sup1} -- 
\ref{fig:sup4}. In addition, the degree of polarization at most epochs varied 
with distance from the core, sometimes exceeding 10\%. 
At some epochs the EVPAs were constant, and at other epochs varied, with distance
from the core. 

\begin{figure*}
\begin{center}
\begin{tabular}{c}
\includegraphics[scale=1.5]{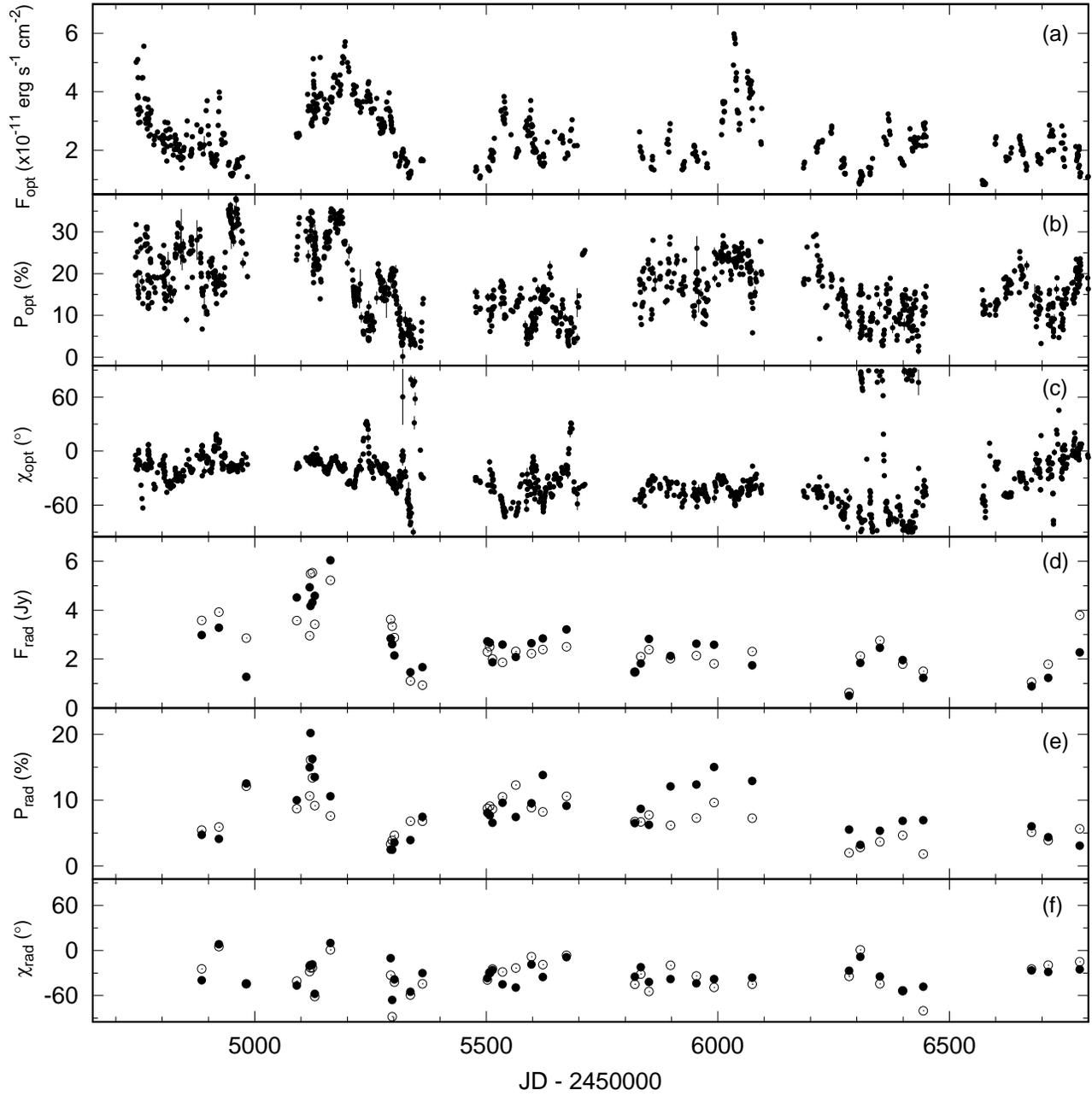}
\end{tabular}
  \caption{Time series of OJ~287; panel (a) - total intensity, panel (b) 
  - degree of polarization, panel (c) - EVPA in the optical band. Panels (d),
  (e), and (f) show the corresponding time series of total intensity, degree 
  of polarization and EVPA, for the measurements at 0.05~mas (open circles) and 
  0.25~mas (filled circles) where the "core" and the stationary feature $S$ are 
  expected to be dominated.}
  \label{fig:timeseries}
\end{center} 
\end{figure*}

Figure~\ref{fig:timeseries} shows a time series of the observed optical 
flux and polarization, as well as radio fluxes and polarization at both 
the core (at 0.05~mas) and $S$ (at 0.25~mas) estimated from individual 
profiles of the intensity, degree of polarization and EVPA. 

\begin{figure*}
\begin{center}
\begin{tabular}{c}
\includegraphics[scale=0.65]{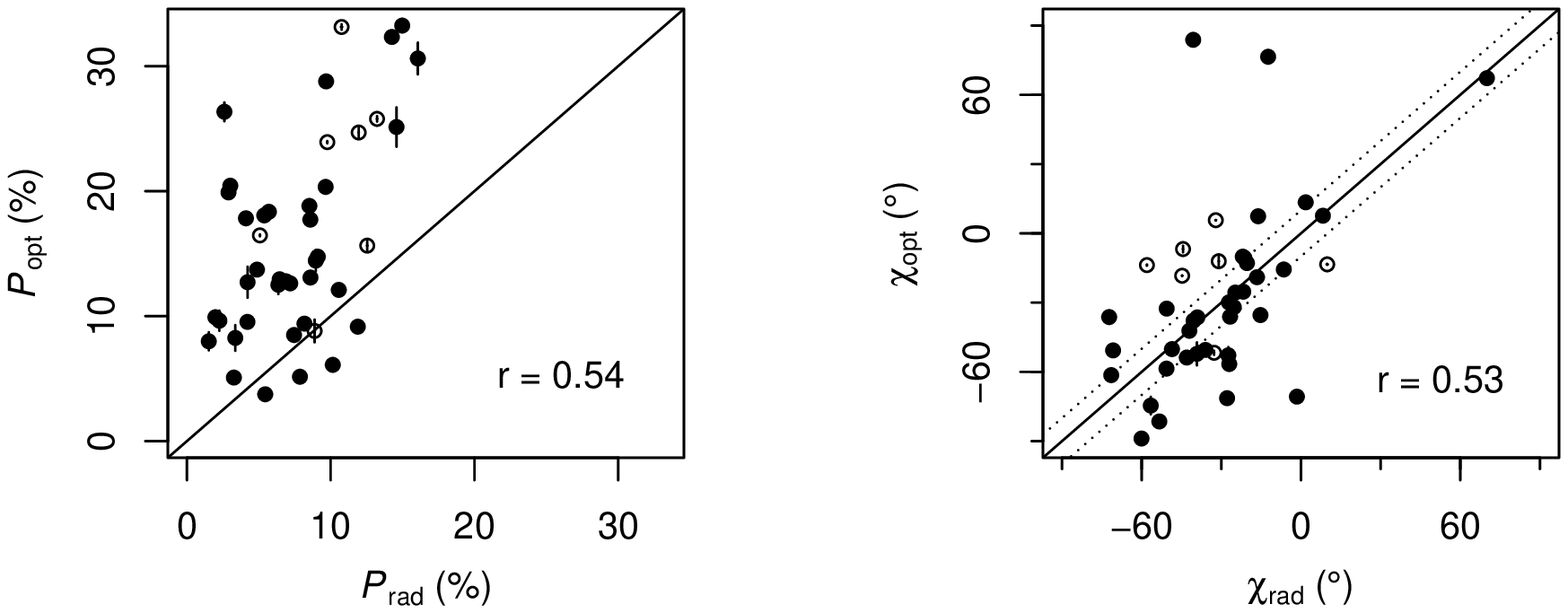}
\end{tabular}
  \caption{Optical vs.\ 7~mm degree of polarization (left) and EVPA 
  (right). The 7~mm values are measured at location of peak total intensity. 
  Solid line corresponds to equal optical and radio values; in right 
  panel, dotted lines encompass EVPA values coinciding within 
  $\pm$10$^{\circ}$. 
  Filled circles: cases when the 7~mm EVPA at one or more positions along 
  the jet trajectory agrees with the optical EVPA within 
  $\pm$10$^{\circ}$; 
  open circles: there are no such positions of similar EVPAs. Correlation 
  coefficients are shown in bottom right of each panel.}
  \label{fig:scatter}
\end{center} 
\end{figure*}

Figure~\ref{fig:scatter} plots the optical versus 7~mm degree of 
polarization at the position of peak total intensity in the VLBA image, 
measured at the same epoch. The degree of optical polarization is mildly 
correlated with that at 7~mm, although the optical values tend to be 
higher. At only 18
epochs out of 43 is the 7~mm EVPA at the location of the total intensity 
peak within $\pm$10$^{\circ}$
of the optical value. There are, however, 18 other epochs 
at which the optical and 7~mm EVPAs coincide within 10$^{\circ}$ 
at positions other than the total intensity peak (see below); these are 
denoted by filled circles outside the dotted lines in the right panel of
Figure~\ref{fig:scatter}. This implies that the primary site of optical 
emission can lie outside the brightest feature (usually the core) in the 
7~mm image. 

Most of the epochs have multiple positions along the jet at which the radio 
and optical EVPAs have similar orientations. One can consider two 
possibilities: (1) the optical emission originates from multiple locations 
along the jet, or (2) the magnetic field is systematically ordered along the 
jet. On the other hand, at seven epochs there are no coinciding positions. 
At some of these epochs the object was in an active state and at others it was
in a quiescent phase. If the optically emitting region is optically thick at the 
radio frequency, the optical emission should not have a relationship 
with the (more extended) radio jet. The other possibility is that the radio 
polarization coming from the optically emitting region is diluted, with its mean
changed by superposition of various polarization vectors associated 
with other regions.

\begin{figure}
\begin{center}
\begin{tabular}{c}
\includegraphics[scale=0.65]{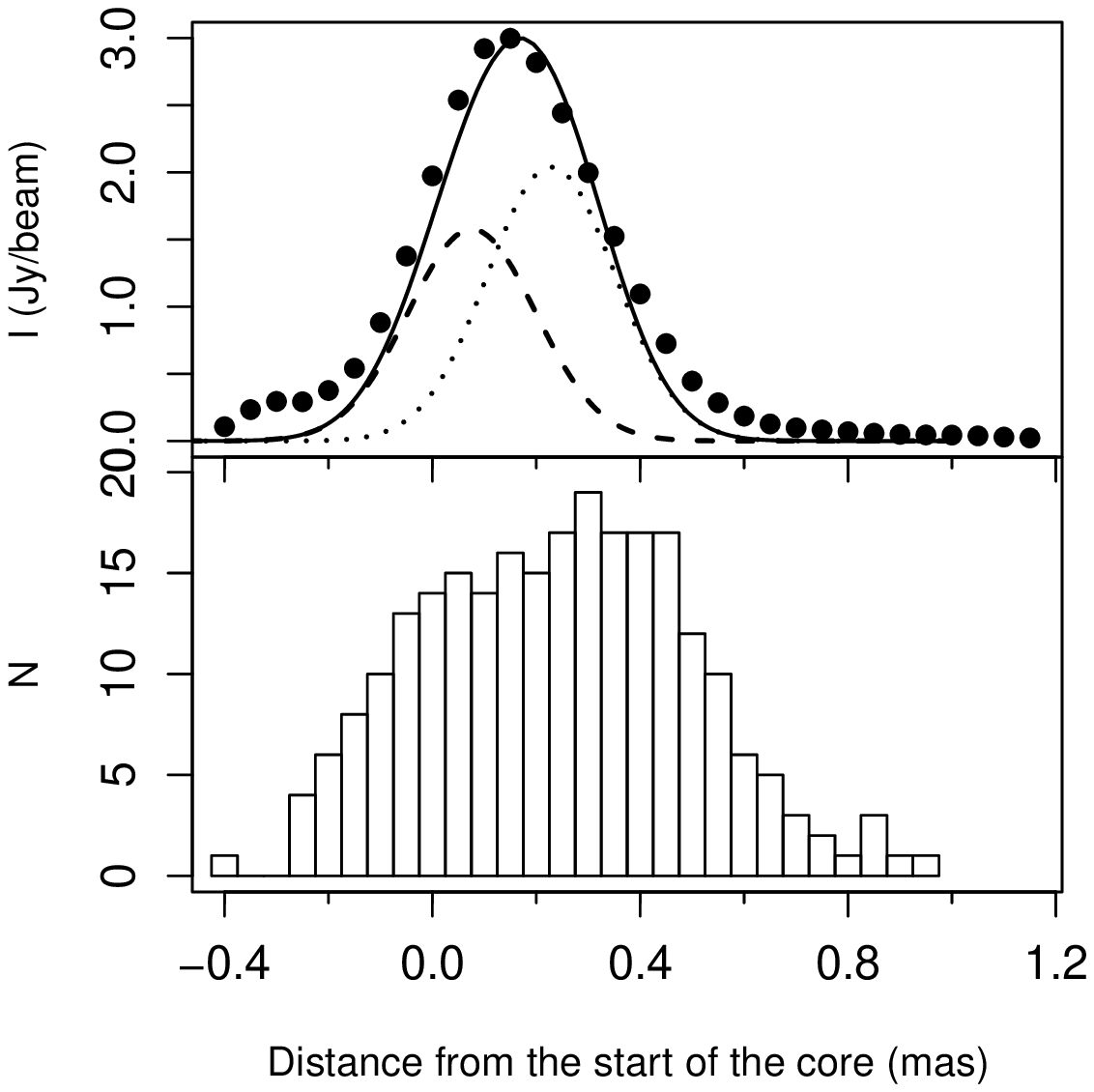}
\end{tabular}
  \caption{{\it Top panel:} Intensity profile along the jet trajectory 
  averaged over all epochs considered here (filled circles). Curves 
  represent the best-fit double point component model (dashed: core, 
  dotted: stationary component $S$, solid: summed intensity). {\it Bottom 
  panel:} Distribution of distances (measured from the upstream boundary 
  of the core) of positions where the optical and 7 mm EVPAs are closest 
  in value (only including those that coincide within 
  $\pm$10$^\circ$).}
  \label{fig:histprofile}
\end{center} 
\end{figure}

The upper panel of Figure~\ref{fig:histprofile} shows the 7~mm intensity 
profile along the jet trajectory, averaged over all 43 epochs. 
We model the profile with two point components to estimate the separate 
intensity profiles of the core and $S$. The model with minimum $\chi^2$ 
value corresponds to average flux densities of the core and $S$ of 1.71 
and 2.14~Jy, respectively, with a separation of 0.20~mas along 
PA $-28^{\circ}$. The solid curve in Figure~\ref{fig:histprofile} shows 
the fit of the model to the data (solid circles). The small deviations 
suggest that both the core and $S$ are slightly resolved rather than 
point-like sources.

As seen in Figure~\ref{fig:histprofile}, the core dominates the intensity 
up to a distance of 0.13 mas, while $S$ dominates from 0.13 to 0.5 mas. 
The bottom panel of the figure demonstrates that the position along the 
jet where the optical and 7 mm EVPAs most closely correspond is sometimes 
associated with the core and other times with $S$. 
Some of the coinciding positions are in regions outside the core 
and $S$. If there is no relationship between the optical and radio EVPAs, the 
possibility of chance coincidence can be estimated from the fraction of EVPA
differences that exceed the threshold of 10$^{\circ}$. This criterion indicates
that the optical and radio EVPAs at 6 out of 43 epochs can be expected to be
similar by chance coincidence. Fewer than 5 coincidences are expected to occur
in the regions outside the core and $S$, but this is too many to determine
whether any optical emission comes from those regions. 
Therefore, the optical emission originates from either the core or 
stationary component $S$ (or both) at the majority of the epochs at which the 
optical and radio EVPAs are aligned.

\subsection{Differences between EVPAs and Jet Position Angle}

\begin{figure}
\begin{center}
\begin{tabular}{c}
\includegraphics[scale=0.6]{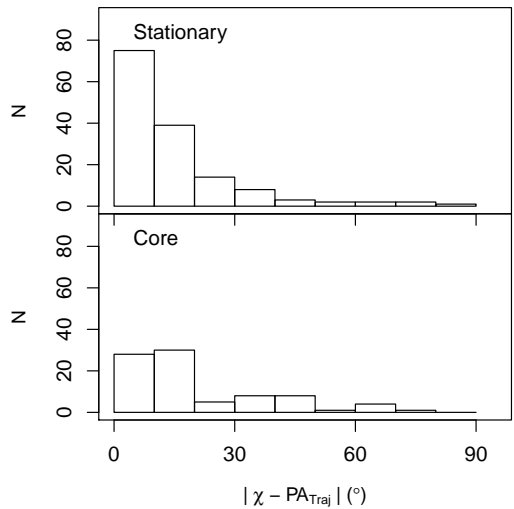}
\end{tabular}
  \caption{Distribution of differences between EVPA at 7~mm and local PA 
  of jet trajectory at epochs when the optical and radio EVPAs match 
  within $20^\circ$ inside either quasi-stationary feature $S$ (upper 
  panel) or (b) the core region (lower panel). (Each 0.05~mas-radius 
  region where the criterion is met is represented.)}
  \label{fig:EVPA-PA}
\end{center} 
\end{figure}

Figure~\ref{fig:EVPA-PA} displays the distributions of differences between 
the PA of the jet trajectory and the EVPAs in cases where the optical and 7 mm 
EVPAs most closely correspond to each other (a) in feature $S$ and (b) 
in the core. Most of the EVPA-PA differences within $S$ are $<30^{\circ}$ 
(skewness score of $2.15$), while those within the core region are more 
broadly distributed (skewness score of $1.34$), although most 
are $<30^{\circ}$.

\citet{2009ApJ...697..985D} measured EVPA values in the 7~mm core in 2006 
that were perpendicular to the previously observed parsec-scale jet 
direction. This caused those authors to propose a spine-plus-sheath 
structure with different magnetic field geometries. However, as cited 
above, the jet changed direction abruptly by $\sim 90^\circ$ shortly 
before those observations. As a consequence, the core EVPA in 2006 was in 
fact nearly parallel to the new jet direction, a condition that continued 
with the observations reported here.

\section{Discussion}

Our finding that optical synchrotron emission from OJ~287 arises in both
the core and stationary feature $S$ agrees with the conclusions by 
\citet{2011ApJ...735L..10A} and \citet{2017A&A...597A..80H}, based on 
timing of multi-waveband variations and motions of knot, that led to the 
conclusion that
$\gamma$-ray and optical flares occur in both regions. This implies that 
both the core and $S$ accelerate electrons to GeV energies. The presence 
of optical emission downstream of the radio ``core'' has also been 
inferred by \citet{2017A&A...598L...1K} from displacements of the optical 
and radio centroids measured by {\it Gaia} and the VLBA.

The magnetic field geometry at the site of optical emission should relate 
to the main process(es) responsible for acceleration of GeV-energy 
electrons. Averaging of the polarization vector both across 
the line-of-sight and within the resolution beam can cause emission from 
turbulent regions --- where second-order Fermi acceleration and magnetic 
reconnections can generate particle distributions extending up to GeV 
energies --- to be weakly polarized, with random (in time) 
orientation of the EVPA. Instead, the polarization is quite strong in 
OJ~287 (Fig. \ref{fig:scatter}) and avoids EVPAs $\gtrsim 50^\circ$ from
the direction of the jet axis (Fig. \ref{fig:EVPA-PA}). 
At some epochs, the optical and radio EVPAs are coincident 
at multiple locations, as in panel (f) of Figure~\ref{fig:Nov}. 
Since the radio polarization vectors are distributed 
along the jet, we cannot distinguish whether the optical emission 
comes from one narrow or multiple more diffuse regions.

The polarization produced by helical fields can be high or low, with EVPA that is 
usually parallel to the jet axis, although transverse EVPAs can arise 
for some viewing angles if the helix pitch angle is low 
\citep{2005MNRAS.360..869L}. Figure~\ref{fig:EVPA-PA} shows that in 
OJ~287 the EVPA is nearly always within $30^\circ$ of the local jet 
direction. Reversals of magnetic field lines leading to reconnections 
can occur in sections of the jet where transverse velocity gradients 
shear magnetic loops into elongated patterns. In this case, the 
polarization would be high, with EVPAs perpendicular to the jet axis, 
contrary to the EVPAs in OJ~287.

A shock wave propagating down the jet, with the shock-normal oriented 
parallel to the jet axis, can cause strong polarization, with EVPA along 
the jet axis \citep{1985ApJ...298..301H}. In OJ~287, however, we have 
identified the main emission as arising in two stationary features, the 
core and $S$. The beam-averaged polarization of turbulent plasma 
crossing a conical \emph{standing} shock can be moderately high 
($\sim$10-20$\%$) with EVPA parallel to the axis 
\citep{1990ApJ...350..536C}. When resolved, the polarization vector has 
a more complex pattern, often roughly radial, centered near the position 
of peak intensity \citep{2006MNRAS.367..851C,2013ApJ...772...14C}. 
\citet{2011ApJ...735L..10A} have proposed that feature $S$ is such a 
standing conical shock.

In either the helical field or conical shock case, the EVPA of the net
polarization can take on various values if there are asymmetries so that 
emission above or below the projected axis is higher than on the other 
side. This can occur, for example, because of turbulent fluctuations 
\citep{2014ApJ...780...87M}.

A less symmetric, nearly planar oblique shock should arise where the jet 
has impinged upon the external medium since it changed direction in 2005. 
However, such a shock would result in an EVPA that is very roughly 
transverse to the jet axis, contrary to the trend in OJ~287.

Particles can be accelerated by a relativistic shock wave in a 
collisionless plasma, with oblique (or conical) shocks favored over 
shocks oriented perpendicular to the jet axis 
\citep[e.g.,][]{2012ApJ...745...63S,2017MNRAS.464.4875B}. This results 
from the magnetic field lying closer to the shock-normal in the rest 
frame of the plasma (after taking into account relativistic aberration) 
and to the flow advecting away from the shock front more slowly than 
when the shock normal is parallel to the jet axis. These features allow 
the relativistic particles to cross the shock front multiple times. The 
accelerated GeV-energy electrons lose energy via radiative cooling as 
they emit optical and (via inverse Compton scattering) $\gamma$-ray 
photons. 
The synchrotron cooling timescale is $\sim 2$~d in the 
observer's frame for a magnetic field strength $\sim0.1$~G and Doppler 
factor $D{\sim}9$ \citep{1995PASJ...47..131T,2010PASJ...62..645S,jor17}, 
hence rapid variability is possible. In this scenario, the optically 
emitting region lies in a thin layer close to the shock front where the 
magnetic field remains compressed \citep{1985ApJ...298..114M}.

The jet of M87 is among the most well-studied jets among active galaxies and 
has a stationary feature known as HST-1. \citet{2016ApJ...832....3A} 
find that the EVPAs in HST-1 are roughly along the jet direction at 
optical wavelengths, but nearly perpendicular at 22~GHz. In OJ~287, the 
similar behavior of the difference between the jet PA and the EVPAs at 
both wavebands suggests that the optical and radio polarization 
patterns are more similar. 

\section{Conclusions}

Our comparison of the total optical polarization and the polarization along the 
jet axis in 7 mm VLBA images confirms the conclusion of 
\citet{2011ApJ...735L..10A} and \citet{2017A&A...597A..80H} that the optical 
emission in OJ~287 arises both in the ``core'' and the bright stationary 
feature $S$ located 0.2 mas downstream of the core. When the optical EVPA 
coincides with that at 7~mm within $S$, the polarization vector 
nearly always lies $<30^\circ$ from the direction of the jet axis. This is 
consistent with $S$ being a standing shock of the type that can accelerate 
electrons to energies high enough to radiate at optical and $\gamma$-ray frequencies.

Since feature $S$ is situated near a bend in the jet trajectory, its high 
intensity --- comparable to and sometimes exceeding that of the core 
\citep{2011ApJ...735L..10A} --- implies that strong optical and $\gamma$-ray 
emission can result from such changes in jet direction. The  ongoing VLBA-BU-BLAZAR monitoring program is well-suited to determine whether this possible 
connection between high-frequency emission and jet bending is common in 
blazars.
\\
\\
The authors thank Dr. Ioannis Myserlis for a critical review of a draft of this manuscript.
MS was supported during this study by a JSPS Postdoctoral Fellowship for Research 
Abroad. The research at Boston University was supported in part by National Science 
Foundation grant AST-1615796 and NASA Fermi Guest Investigator Program grants 
NNX14AQ58G and 80NSSC17K0649. The St.Petersburg University team acknowledges
support from Russian Science Foundation grant 17-12-01029.
IA acknowledges support by a Ram\'on y Cajal grant of the Ministerio de Econom\'ia y 
Competitividad (MINECO) of Spain. The research at the IAA--CSIC was supported in 
part by the MINECO through grants AYA2016-80889-P, AYA2013-40825-P, and 
AYA2010-14844, and by the regional government of Andaluc\'{i}a through grant 
P09-FQM-4784. Calar Alto Observatory is jointly operated by the Max-Planck-Institut 
f\"ur Astronomie and the IAA-CSIC. The VLBA is an instrument of the Long Baseline 
Observatory (LBO). The LBO is a facility of the National Science Foundation operated 
under cooperative agreement by Associated Universities, Inc. Data from the Steward 
Observatory spectropolarimetric monitoring project were used. This program was 
supported by Fermi Guest Investigator grants NNX08AW56G, NNX09AU10G, 
NNX12AO93G, and NNX15AU81G.

\appendix \label{appendix}

\begin{figure}
\begin{center}
\begin{tabular}{c}
\includegraphics[width=18cm]{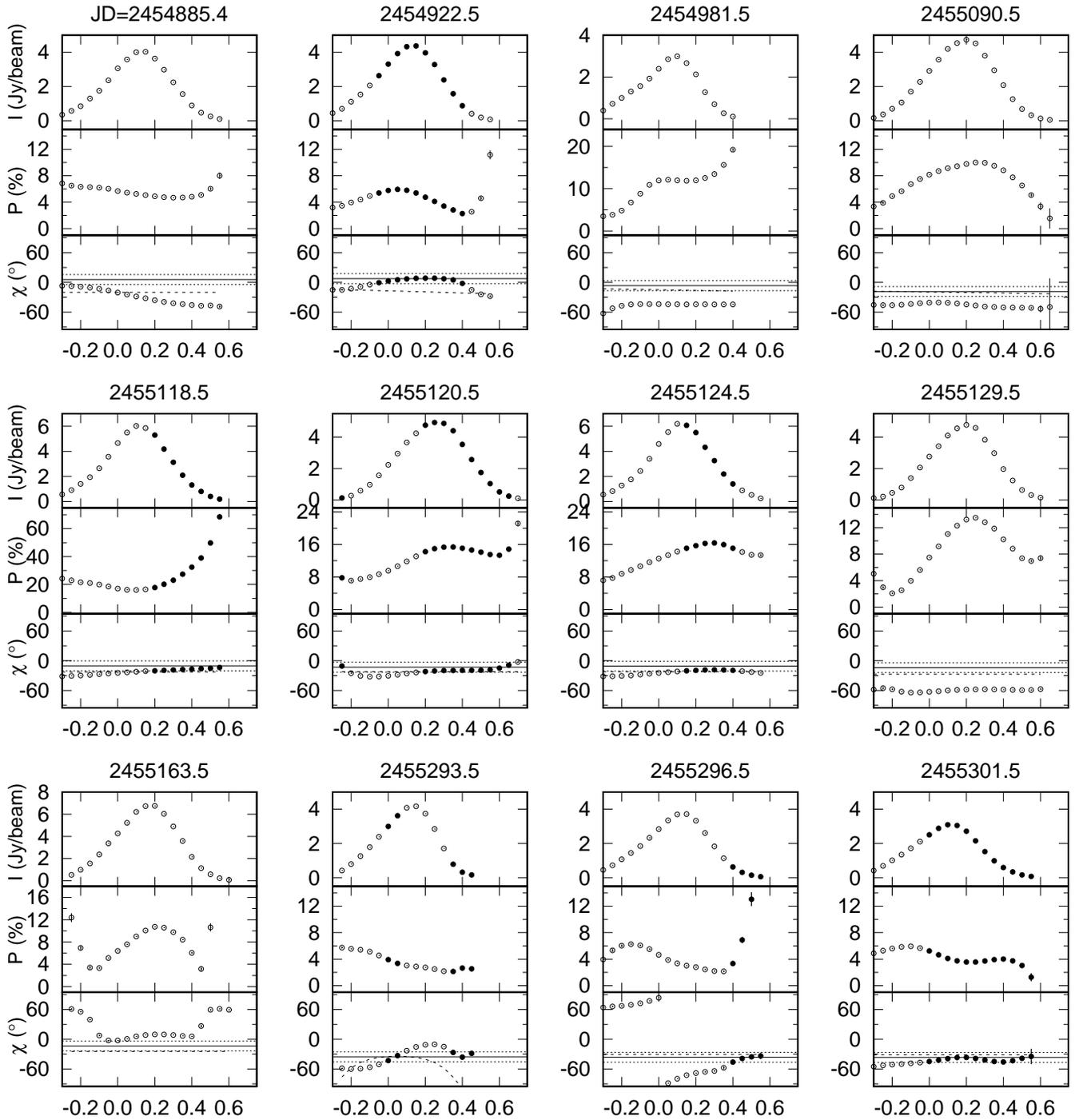}
\end{tabular}
  \caption{Profiles of intensites, degrees of polarization and EVPAs along the 
  jet trajectory from the epoch of February 2009 to April 2010. }
  \label{fig:sup1}
\end{center} 
\end{figure}

\begin{figure}
\begin{center}
\begin{tabular}{c}
\includegraphics[width=18cm]{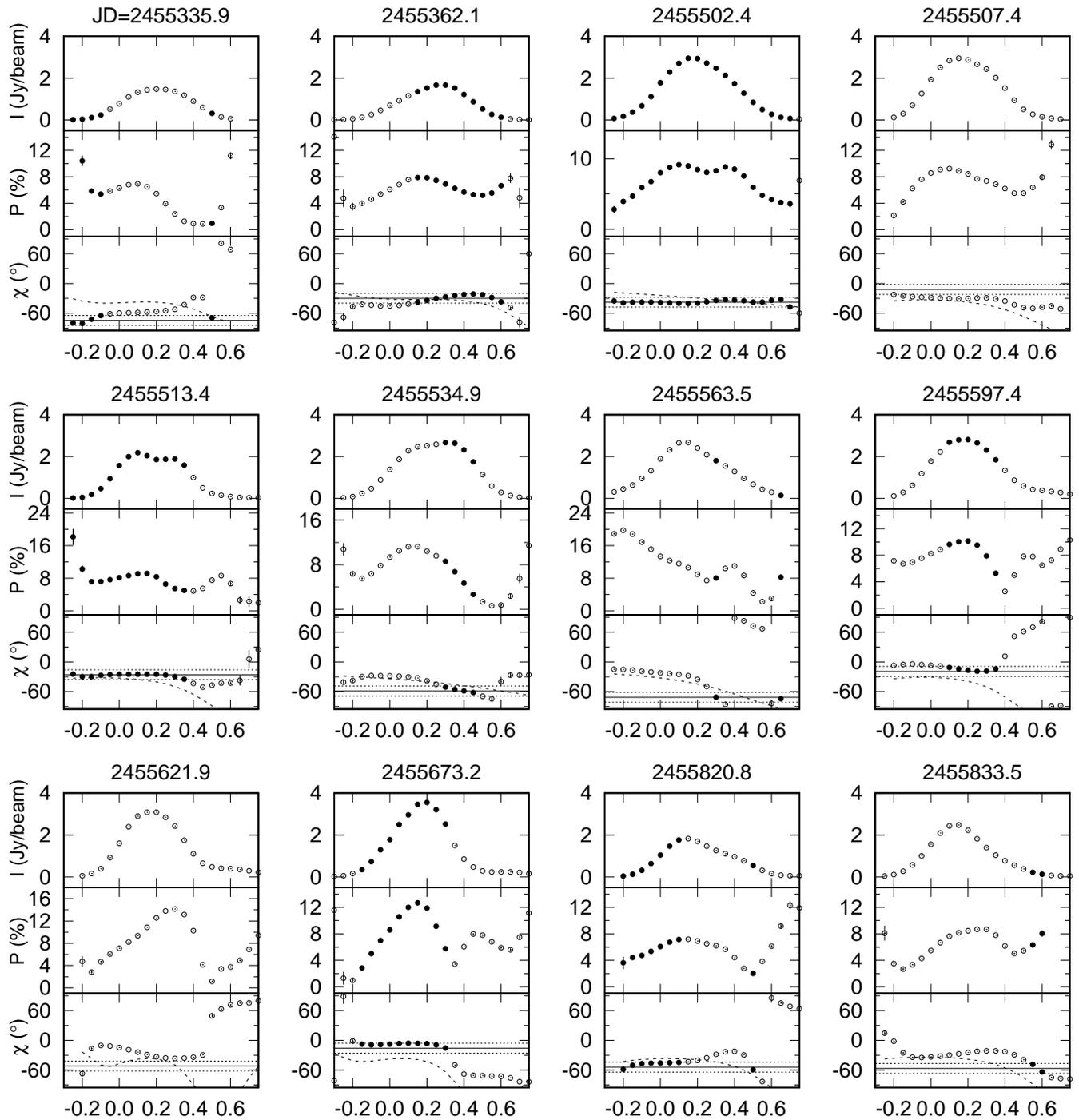}
\end{tabular}
  \caption{Epochs of May 2010 to September 2011.}
  \label{fig:sup2}
\end{center} 
\end{figure}

\begin{figure}
\begin{center}
\begin{tabular}{c}
\includegraphics[width=18cm]{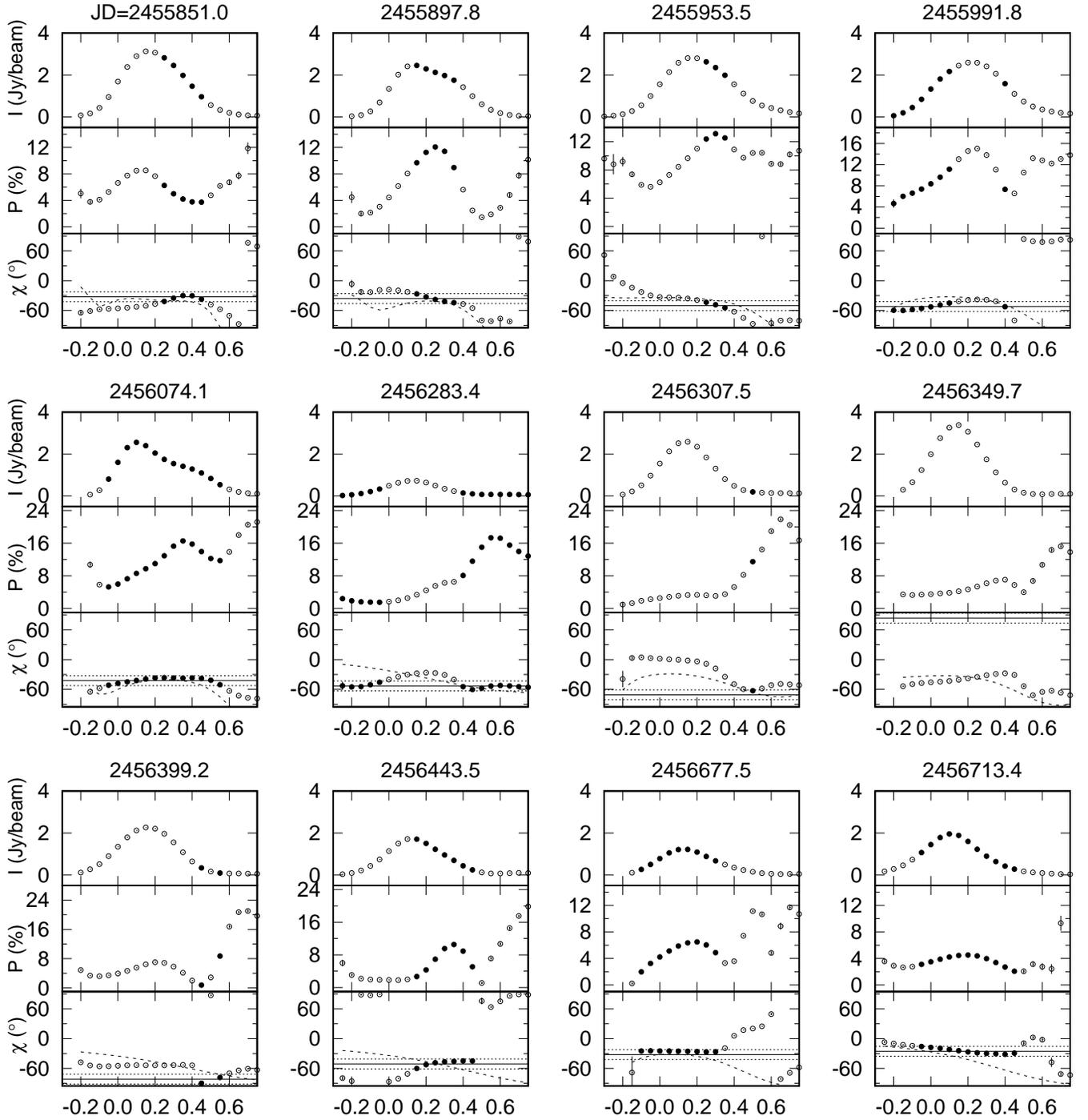}
\end{tabular}
  \caption{Epochs of October 2011 to February 2014.}
  \label{fig:sup3}
\end{center} 
\end{figure}

\begin{figure}
\begin{center}
\begin{tabular}{c}
\includegraphics[width=18cm]{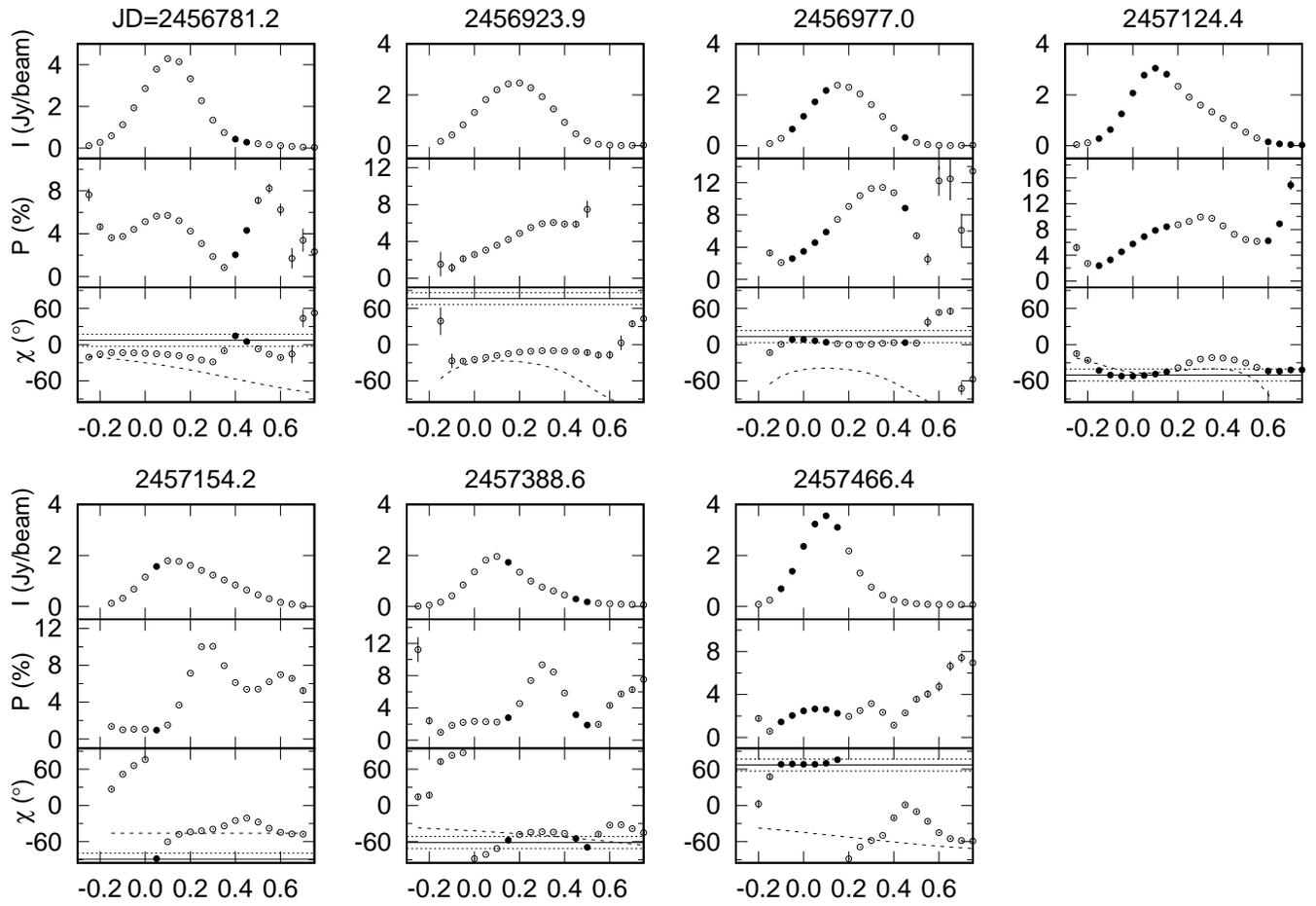}
\end{tabular}
  \caption{Epochs of May 2014 to March 2016.}
  \label{fig:sup4}
\end{center} 
\end{figure}

Figures~\ref{fig:sup1}--\ref{fig:sup4} present profiles of intensities, degrees 
of polarization and EVPAs along the jet trajectory at 43 epochs. Displayed from 
top to bottom at each epoch are the intensity, degree of polarization, and EVPA. 
In the bottom panel, the dashed line indicates the PA of the jet trajectory, 
while the solid and dotted lines mark the optical EVPA and the thresholds of 
$\pm$10$^\circ$, respectively.

\end{document}